\documentclass[aps,prl,twocolumn,showpacs,preprintnumbers,amsmath,amssymb,superscriptaddress]{revtex4}

\usepackage{dcolumn}
\usepackage{bm}
\usepackage{ulem} 
\usepackage[dvips]{graphicx}
\usepackage{color}
\usepackage[dvips]{graphicx}

\begin{document}
\def\Journal#1#2#3#4{{#1} {\bf #2}, #3 (#4)}
\def\PLB{{Phys. Lett.}  B}
\def\PRL{{Phys. Rev. Lett.}}
\def\PRC{{Phys. Rev.} C}
\def\PRD{{Phys. Rev.} D}
\def\NIMA{{Nucl. Instr. Meth.} A}
\def\EPJ{{Eur. Phys. J.} C}


\title{Measurement of the Bottom contribution to non-photonic electron production in $p+p$ collisions at $\sqrt{s} $=200 GeV}

\affiliation{Argonne National Laboratory, Argonne, Illinois 60439, USA}
\affiliation{University of Birmingham, Birmingham, United Kingdom}
\affiliation{Brookhaven National Laboratory, Upton, New York 11973, USA}
\affiliation{University of California, Berkeley, California 94720, USA}
\affiliation{University of California, Davis, California 95616, USA}
\affiliation{University of California, Los Angeles, California 90095, USA}
\affiliation{Universidade Estadual de Campinas, Sao Paulo, Brazil}
\affiliation{University of Illinois at Chicago, Chicago, Illinois 60607, USA}
\affiliation{Creighton University, Omaha, Nebraska 68178, USA}
\affiliation{Czech Technical University in Prague, FNSPE, Prague, 115 19, Czech Republic}
\affiliation{Nuclear Physics Institute AS CR, 250 68 \v{R}e\v{z}/Prague, Czech Republic}
\affiliation{University of Frankfurt, Frankfurt, Germany}
\affiliation{Institute of Physics, Bhubaneswar 751005, India}
\affiliation{Indian Institute of Technology, Mumbai, India}
\affiliation{Indiana University, Bloomington, Indiana 47408, USA}
\affiliation{Alikhanov Institute for Theoretical and Experimental Physics, Moscow, Russia}
\affiliation{University of Jammu, Jammu 180001, India}
\affiliation{Joint Institute for Nuclear Research, Dubna, 141 980, Russia}
\affiliation{Kent State University, Kent, Ohio 44242, USA}
\affiliation{University of Kentucky, Lexington, Kentucky, 40506-0055, USA}
\affiliation{Institute of Modern Physics, Lanzhou, China}
\affiliation{Lawrence Berkeley National Laboratory, Berkeley, California 94720, USA}
\affiliation{Massachusetts Institute of Technology, Cambridge, MA 02139-4307, USA}
\affiliation{Max-Planck-Institut f\"ur Physik, Munich, Germany}
\affiliation{Michigan State University, East Lansing, Michigan 48824, USA}
\affiliation{Moscow Engineering Physics Institute, Moscow Russia}
\affiliation{City College of New York, New York City, New York 10031, USA}
\affiliation{NIKHEF and Utrecht University, Amsterdam, The Netherlands}
\affiliation{Ohio State University, Columbus, Ohio 43210, USA}
\affiliation{Old Dominion University, Norfolk, VA, 23529, USA}
\affiliation{Panjab University, Chandigarh 160014, India}
\affiliation{Pennsylvania State University, University Park, Pennsylvania 16802, USA}
\affiliation{Institute of High Energy Physics, Protvino, Russia}
\affiliation{Purdue University, West Lafayette, Indiana 47907, USA}
\affiliation{Pusan National University, Pusan, Republic of Korea}
\affiliation{University of Rajasthan, Jaipur 302004, India}
\affiliation{Rice University, Houston, Texas 77251, USA}
\affiliation{Universidade de Sao Paulo, Sao Paulo, Brazil}
\affiliation{University of Science \& Technology of China, Hefei 230026, China}
\affiliation{Shandong University, Jinan, Shandong 250100, China}
\affiliation{Shanghai Institute of Applied Physics, Shanghai 201800, China}
\affiliation{SUBATECH, Nantes, France}
\affiliation{Texas A\&M University, College Station, Texas 77843, USA}
\affiliation{University of Texas, Austin, Texas 78712, USA}
\affiliation{Tsinghua University, Beijing 100084, China}
\affiliation{United States Naval Academy, Annapolis, MD 21402, USA}
\affiliation{Valparaiso University, Valparaiso, Indiana 46383, USA}
\affiliation{Variable Energy Cyclotron Centre, Kolkata 700064, India}
\affiliation{Warsaw University of Technology, Warsaw, Poland}
\affiliation{University of Washington, Seattle, Washington 98195, USA}
\affiliation{Wayne State University, Detroit, Michigan 48201, USA}
\affiliation{Institute of Particle Physics, CCNU (HZNU), Wuhan 430079, China}
\affiliation{Yale University, New Haven, Connecticut 06520, USA}
\affiliation{University of Zagreb, Zagreb, HR-10002, Croatia}

\author{M.~M.~Aggarwal}\affiliation{Panjab University, Chandigarh 160014, India}
\author{Z.~Ahammed}\affiliation{Lawrence Berkeley National Laboratory, Berkeley, California 94720, USA}
\author{A.~V.~Alakhverdyants}\affiliation{Joint Institute for Nuclear Research, Dubna, 141 980, Russia}
\author{I.~Alekseev~~}\affiliation{Alikhanov Institute for Theoretical and Experimental Physics, Moscow, Russia}
\author{J.~Alford}\affiliation{Kent State University, Kent, Ohio 44242, USA}
\author{B.~D.~Anderson}\affiliation{Kent State University, Kent, Ohio 44242, USA}
\author{DanielAnson}\affiliation{Ohio State University, Columbus, Ohio 43210, USA}
\author{D.~Arkhipkin}\affiliation{Brookhaven National Laboratory, Upton, New York 11973, USA}
\author{G.~S.~Averichev}\affiliation{Joint Institute for Nuclear Research, Dubna, 141 980, Russia}
\author{J.~Balewski}\affiliation{Massachusetts Institute of Technology, Cambridge, MA 02139-4307, USA}
% ...

\collaboration{STAR Collaboration}\noaffiliation

\date{\today}

\begin{abstract}
The contribution of $B$ meson decays to non-photonic electrons, which are mainly produced by the semi-leptonic decays of heavy flavor mesons,
in $p+p$ collisions at $\sqrt{s} =$ 200 GeV has been measured using azimuthal correlations between non-photonic electrons and hadrons.
The extracted $B$ decay contribution is approximately 50\% at a transverse momentum of $p_{T} \geq 5$ GeV/$c$. 
These measurements constrain the nuclear modification factor for electrons from $B$ and $D$ meson decays.
The result indicates that $B$ meson production in heavy ion collisions is also suppressed at high $p_{T}$.
\end{abstract}


\maketitle

The suppression of non-photonic electron yields from semi-leptonic decays of $D$ and $B$ mesons for $p_{T}$ up to 9 GeV/$c$ in central Au+Au collisions at Relativistic Heavy Ion Collider (RHIC) 
has been observed to be large \cite{PHENIX_e}, and similar to that of light quark hadrons \cite{STAR_RAA}. 
Due to the dead cone effect, heavy quarks were expected
to lose less energy than light quarks \cite{deadcone} if the dominant energy loss mechanism is gluon radiation \cite{rad}.
Various models have been proposed to explain the large suppression of non-photonic electron yields \cite{eloss_1,eloss_dessociate,eloss_reso}.
Theoretical calculations of the non-photonic electron suppression crucially depend on the $B/D$ ratios because the amount of radiative energy loss depends on the quark mass.
Measuring the bottom quark contribution to non-photonic electron yields in $p+p$ collisions is therefore important in order to understand 
the production of heavy quarks
and to provide a baseline for the energy loss measurement of heavy quarks in the hot and dense medium produced in central Au+Au collisions.

In this paper, we report  a determination of the relative contribution from $B$ decays to 
non-photonic electron yields ($r_{B}$) by measuring  the azimuthal correlations between non-photonic electrons and charged hadrons ($e_{\textrm{non}\gamma} \textrm{-} h$),
and between non-photonic electrons and $D^0$ mesons ($e_{\textrm{non}\gamma}\mathchar`-D^{0}$)
in $p+p$ collisions at $\sqrt{s} = 200$ GeV by the STAR experiment at RHIC.
We fit the experimental $e_{\textrm{non}\gamma} \mathchar`- h$ correlations using a combination of PYTHIA calculations \cite{PYTHIA} for $D$  and $B$ meson decays and extract $r_{B}$ as a function of $p_{T}$ ($2.5 <p_{T} <9.5$ GeV/$c$). 
An independent measurement of the $r_{B}$ is obtained from $e_{\textrm{non}\gamma} \mathchar`- D^{0}$ correlations, by selecting the
charge combinations $e^{-} \mathchar`- D^{0}(\rightarrow K^{-})$ and $e^{+} \mathchar`- \overline{D}^{0}(\rightarrow K^{+})$, which provide
relatively pure samples of $B$ decays and charm pairs on near and away-side ($\Delta\phi\sim\pi$) \cite{eD0sim}.
The combined measurements of the $B$ decay contribution and of the nuclear modification factor ($R_{AA}$) for heavy-flavor decay electrons in Au+Au collisions
constrain the value of the $R_{AA}$ for electrons from $B$ meson decays.

The $p+p$ data used in this analysis were taken by the STAR experiment  \cite{STAR} during the 2005 and 2006 RHIC runs.
The main detectors for this analysis are the Time Projection Chamber (TPC) and the Barrel Electromagnetic Calorimeter (BEMC).
The BEMC has a Shower Maximum Detector (SMD): proportional gas chambers with strip readout at a 
depth of $\sim$ 5 radiation lengths ($X_{0}$) designed to measure shower shapes and positions.
The acceptance for electrons in pseudorapidity and azimuth is $|\eta|<0.7$ ($0<\eta<0.7$ in the 2005 run) and $0<\phi<2\pi$.
The BEMC also serves as a trigger detector for high $p_T$ electrons or photons,
where single-tower transverse energy thresholds of 2.6~GeV and 5.4~GeV were used.
The total sampled luminosity was 11.3 pb$^{-1}$ (0.65 pb$^{-1}$) for the 5.4 GeV (2.6 GeV) trigger threshold.
We used triggered events with primary vertices located within 35 cm of the TPC's geometrical center along the beam direction.

Electrons were identified by measuring ionization energy loss ($dE/dx$) and track momentum ($p$) from TPC,
the energy ($E$)  deposition in the BEMC, and the shower profile in the SMD. 
A significant fraction of the hadron background was rejected by
selecting tracks with a measured $dE/dx$ in the TPC between $-1$ and $+3$ standard deviations from the expected mean $dE/dx$ for electrons.
Based on calibrations of the SMD response to electrons and hadrons,
tracks whose shower projection occupies more than 1 strip in both $\phi$ and $\eta$ SMD planes were selected as electron candidates.
We required the energy-to-momentum ratio to be in the range $0.3< p/E <1.5$.
The hadron contamination in the electron sample after applying these cuts is
$\sim2 \%$ up to 5 GeV/$c$, increasing to $\sim10 \%$ at 9 GeV/$c$. 

The electron sample has two components:
(1) non-photonic  electrons and 
(2) photonic electrons - those from photon conversion in the detector material between the interaction point and 
the TPC and Dalitz decays, mainly from $\pi^{0}$.
Photonic electrons were identified by pairing electrons with oppositely charged partner tracks, 
determining the conversion or decay vertex, and calculating the invariant mass of the $e^{+}e^{-}$ pair $M_{e^{+}e^{-}}$ ~\cite{2Dmass}.
To improve the invariant mass resolution, the so-called 2-D invariant mass was calculated using only $p_T$ and $p_Z$, which is equivalent to setting the opening angle in the transverse plane to zero \cite{2Dmass}.
Monte Carlo simulations indicate that the cut of 0.1
GeV/$c^2$ removes almost all photon conversion candidates for which the decay partner is reconstructed in the TPC.
The efficiencies for photonic electron reconstruction ($\epsilon_{e_{\gamma}}$) range from $65\%$ at 3.0 GeV/$c$ to $80\%$ at 8.0 GeV/$c$,
as determined from GEANT simulations.
\begin{figure}
\includegraphics[width=1.0\linewidth,height = 5.5 cm,clip]{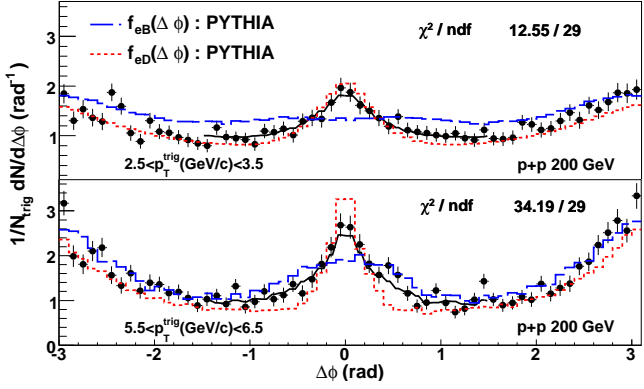}
\caption{ (color online) Distributions of the azimuthal angle between non-photonic electrons and charged hadrons
normalized per non-photonic electron trigger.
The trigger electron has (top) 2.5 $<p_{T}<$ 3.5 GeV/$c$ and (bottom) 5.5 $<p_{T}<$ 6.5 GeV/$c$.
The curves represent PYTHIA calculations for $D$ (dotted curve) and $B$ (dashed curve) decays.
The fit result is shown as the black solid curve.}
\label{fig:fig1}
\end{figure}
For the  $e_{\textrm{non}\gamma} \mathchar`- h$ analysis,
we first removed the electrons that have an opposite-sign partner such that $M_{e^{+}e^{-}}<0.1$ GeV/$c^{2}$ from the inclusive electron sample.
The remaining electrons form the ``semi-inclusive" electron sample.
The non-photonic electron yields can be expressed as,
\begin{equation}\label{bla0}
N_{e_{\textrm{non}\gamma}} = N_{{e}_{\textrm{semi}}} + N_{e_{\textrm{like}}} - N_{e_{\gamma}^{\textrm{not-reco}}}-N_{h}.
\end{equation}
$N_{{e}_{\textrm{semi}}}$ is the number of semi-inclusive electrons.
$ N_{e_{\gamma}^{\textrm{not-reco}}}$ represents the number of photonic electrons which are not reconstructed by the invariant mass method 
and is defined as: $(1/\epsilon_{e_{\gamma}}-1)(N_{e_{\textrm{unlike}}} - N_{e_{\textrm{like}}})$.
$N_{e_{\textbf{like}}}$ is the number of non-photonic electrons  that were rejected by the conversion cuts because they happened to form a pair with a random track
which is determined using like-sign pairs.
$N_{h}$ is the remaining background from hadron contamination in the electron sample.
Other weak decay contributions such as $K_{e3}$  are negligible due to their long $c{\tau}$,
and charmed baryons (mostly $\Lambda_{c}$) is expected to be very small contribution since the baryon yield 
is small compared to the meson yield ($\Lambda_{c}/D^{0} \sim 0.1$ in PYTHIA) and the branching ratio for semi-leptonic decays is smaller for baryons than mesons.
The $e_{\textrm{non}\gamma} \mathchar`-h$ azimuthal distributions were calculated as
\begin{eqnarray}\label{bla1}
\frac{dN^{e_{\textrm{non}\gamma} \mathchar`-h}}{d(\Delta\phi)} &
=& \frac{dN^{e_{\textrm{semi}} \mathchar`-h}}{d(\Delta\phi)} + 
\frac{dN^{e_{\textrm{like}} \mathchar`-h}}{d(\Delta\phi)}-\frac{dN^{e^{\textrm{not-reco}}_{\gamma} \mathchar`-h}}{d(\Delta\phi)}  \nonumber \\
                                                                               &-&\frac{dN^{h\mathchar`-h}}{d(\Delta\phi)},                                                                    
\end{eqnarray}
where each term is normalised to be per non-photonic electron trigger.
Each angle-difference distribution on the right-hand side of Eq.~\eqref{bla1} was experimentally determined.
The distribution $dN^{e^{\textrm{not-reco}}_{\gamma} \mathchar`-h}/d(\Delta\phi)$ was constructed from $dN^{e^{\textrm{reco}}_{\gamma} \mathchar`-h}/d(\Delta\phi)$ by removing the conversion partner
to account for the fact that the partner electron is not reconstructed.

Figure~\ref{fig:fig1} shows $dN^{e_{\textrm{non}\gamma} \mathchar`-h}/d(\Delta\phi)$ per trigger for non-photonic electrons for two different trigger $p_{T}$ selections.
Associated particles were required to have $p_{T} >0.3$ GeV/$c$ and $|\eta|<1.05$.
The dotted (dashed) line in the figure represents a PYTHIA version 6.22 calculation of the azimuthal correlations between electrons from $D$ ($B$) meson decay and charged hadrons ($f_{e_{D}}(\Delta \phi)$, $f_{e_{B}}(\Delta\phi)$) \cite{e_h}.
PYTHIA was tuned to reproduce the shapes of $p_{T}$ distributions for $D$ mesons measured by STAR \cite{starD,e_h}.
The PYTHIA calculation shows that the near-side peak for $f_{e_{B}}(\Delta\phi)$ is broader than that for $f_{e_{D}}(\Delta\phi)$.
These shapes are dominated by decay kinematics.
The fragmentation function does not affect the shape in a significant way.
The fraction of non-photonic electrons from $B$ meson decay can be determined by fitting the near-side distribution function ($|\Delta \phi| < $ 1.5):
\begin{equation}\label{fitfunc}
\frac{1}{N^{\textrm{non}\gamma}_{\textrm{trig}}}\frac{dN^{e_{\textrm{non}\gamma \mathchar`-h}}}{d(\Delta\phi)} = r_{B}f_{e_{B}}(\Delta\phi) + (1-r_{B})f_{e_{D}}(\Delta\phi),
\end{equation}
where $r_{B}$ is the ratio of electrons from $B$ meson decay to the total non-photonic electron yield,
$r_{B} = N_{e_{B}}/(N_{e_{B}}+N_{e_{D}})=N_{e_{B}}/N_{e_{\textrm{non}\gamma}}$.

An independent measurement of $r_{B}$ was performed using $e_{\textrm{non}\gamma} \mathchar`- D^{0}$ correlations.
$D^0$ mesons  were reconstructed via their hadronic decay $D^0 \rightarrow K^- \pi^+$ ($\mathcal{B}$ = 3.89$\%$) by calculating the invariant mass of all oppositely charged TPC tracks in the same event.
In this analysis, only events with a non-photonic electron trigger were used for $D^0$ reconstruction. 
Furthermore, the kaon candidates were required to have the same charge sign as the non-photonic electrons \cite{eD0sim}.
The combinatorial background of random pairs was evaluated by combining all charged tracks with the same charge sign from the same event.  
The requirement of a non-photonic electron trigger suppresses the combinatorial background, 
yielding a signal ($S$)-to-background ($B$) ratio of about $14\%$ and a signal significance ($S/\sqrt{S+B}$) of  $\sim4.6$. 

\begin{figure}
\includegraphics[width=1.0\linewidth,height = 5. cm,clip]{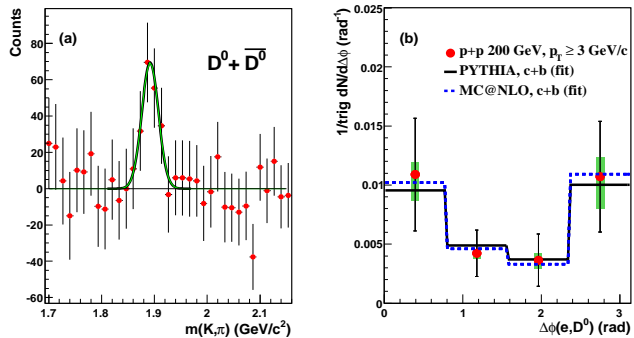}
\caption{(a) Background-subtracted invariant mass distribution of $K\pi$ pairs requiring at least one non-photonic electron trigger in the event. The solid line is a Gaussian fit to the data near the peak region. 
(b) Distribution of the azimuthal angle between non-photonic electron (positron) trigger particles and $D^{0} (\overline{D}^{0})$. 
The solid (dashed) line is a fit of the correlation function from PYTHIA (MC@NLO) simulations to the data points.}
\label{fig:fig2}
\end{figure}

Figure ~\ref{fig:fig2} (a) shows the background subtracted $\pi$-$K$ invariant mass distribution. 
The peak position and width were determined using a Gaussian fit to the data. 
The $K\pi$ invariant mass distribution was obtained for different $\Delta \phi$ bins with respect to the trigger electron, 
and the yield of the associated $D^0$ mesons was taken as the area underneath the Gaussian fit to the signal. 
Figure~\ref{fig:fig2} (b) shows the azimuthal correlation of $e_{\textrm{non}\gamma} \mathchar`-D^{0}$, which exhibits near- and away-side correlation peaks with similar yields.
The results are fitted with the correlation functions for charm and bottom production from PYTHIA and MC@NLO simulations having the relative $B$ contribution as a free parameter \cite{eD0sim}.
The observed away-side correlation peak can be attributed to prompt charm pair production ($\sim75\%$) and $B$ decays ($\sim25\%$), whereas the contributions to the near-side peak are mainly from $B$ decays. 
We determined $r_{B}$ by fitting the measured $e_{\textrm{non}\gamma} \mathchar`-D^{0}$ correlation with PYTHIA and MC@NLO
and used the average of the two fits for the final value.

\begin{figure}[t]
\includegraphics[width = 1.0\linewidth,height = 5.4 cm]{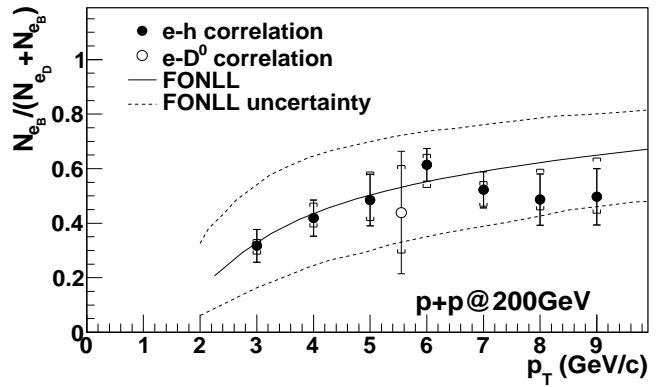}
\caption{Transverse momentum dependence of the relative contribution from $B$ mesons ($r_{B}$) to the non-photonic electron yields. 
Error bars are statistical and brackets are systematic uncertainties.
The solid curve is the FONLL calculation \cite{FONLL}. Theoretical uncertainties are indicated by the dashed curves.
}
\label{fig:fig3}
\end{figure}

Figure~\ref{fig:fig3} shows $r_{B} =N_{e_{B}}/(N_{e_{B}}+N_{e_{D}})$
extracted from $e_{\textrm{non}\gamma} \mathchar`-h$ correlations (filled circles)
and $e_{\textrm{non}\gamma} \mathchar`-D^{0}$ correlations (open circle)  as a function of $p_{T}$.
The vertical lines represent the statistical errors and the systematic uncertainties are shown as brackets.
The systematic uncertainties due to electron identification ($\sim 7 \%$ ), photonic electron
rejection ($\sim 6 \%$),
the fit range ($\sim 10 \%$) and the normalization of the azimuthal distribution ($\sim 10 \%$),
PYTHIA and MC@NLO predictions for $e_{non\gamma}-D^{0}$ ($\sim 5 \%$), 
and the $D^{0}$ signal extraction were estimated by varying the
associated cut parameters and adding the individual contributions in quadrature.
$r_{B}$ increases with electron $p_{T}$ and reaches approximately 0.5 ($N_{e_{B}}/N_{e_{D}} \sim 1$) around $p_{T} = 5$ GeV/$c$.
$r_{B}$ from  the $e_{\textrm{non}\gamma} \mathchar`-D^{0}$ correlation measurement at $p_{T}$ $\sim$ 5.5 GeV/c 
is consistent with $r_{B}$ from $e_{\textrm{non}\gamma}\mathchar`-h$ correlations.
The curve in the figure is $r_{B}$ from a FONLL pQCD calculation
including theoretical uncertainties  \cite{FONLL}.
Similar ratios at $ 2 <p_{T}< 7$ GeV/$c$ using a different method have also been reported \cite{bphenix}.
$J/\psi$ di-electron decays can also contribute to non-photonic electrons and STAR measurement of
$J/\psi$ at high $p_{T}$ indicates that $J/\psi$ decays could contribute nearly 10~\% around $p_{T}$ 5 GeV/$c$. 
The estimated effect of the electrons from $J/\psi$ decays on
$r_{B}$ is a few percent, much smaller than the current statistical and
systematic uncertainties, and no correction was applied to our data.

Next, we explore the implications of the measured $r_{B}$ for the $R_{AA}$ of electrons from $B$ meson decay in heavy ion collisions.
The $R_{AA}$ for heavy flavor non-photonic electrons ($R^{\textrm{HF}}_{AA}$ ) is given by 
\begin{equation}
R_{AA}^{\textrm{\textrm{HF}}} =  (1-r_{B})R_{AA}^{e_{D}} + r_{B}R_{AA}^{e_{B}},
\label{eq:raa_corr}
\end{equation}
where $R_{AA}^{e_{D}}$ ($R_{AA}^{e_{B}}$) is the $R_{AA}$ for electrons from $D$ ($B$) mesons. 
From Eq.~(\ref{eq:raa_corr}), $R_{AA}^{e_{D}}$ and $R_{AA}^{e_{B}}$ are related by the $B$ decay contribution to the non-photonic electron yields ($r_{B}$) in $p+p$ collisions. 
We have taken the $R^{\textrm{HF}}_{AA}$ measurement from PHENIX \cite{PHENIX_e_new, snpe} and
fit the $R^{\textrm{HF}}_{AA}$ above $p_{T}> 5$ GeV/$c$ to a constant value and obtained: $R_{AA} =
0.167^{+0.0562}_{-0.0485}~\textrm{(stat)} ^{+0.0512} _{- 0.0815}~\textrm{(syst)}\pm0.0117~\textrm{(norm)}$, where the
statistical and systematic errors are evaluated from
weighted average over these $p_T> 5$ GeV/$c$ points. 
We also calculate the weighted mean $r_B$ value for $p_{T}> 5$
GeV/$c$ including statistical and systematic errors from our measurement: $r_{B}
= 0.54 \pm 0.0349~\textrm{(sta.)} \pm 0.0666~\textrm{(sys.)}$. Then using Eq.~4 we calculate a likelihood
distribution for $R_{AA}^{e_{B}}$ as a function of $R_{AA}^{e_{D}}$ and the results are shown
in Fig. 4. The most probable values for the $R^{e_{D}}_{AA}$ and $R^{e_{B}}_{AA}$ correlation
are shown by the line with open circles and the 90\% Confidence Limit curves
are represented by dashed lines.
This result indicates that $B$ meson yields are suppressed at high $p_{T}$
in heavy ion collisions 
presumably due to energy loss of the $b$ quark in the dense medium \cite{eloss_1} or of the heavy flavor hadrons due to dissociation \cite{eloss_dessociate} or elastic scattering ~\cite{eloss_reso}.
Our conclusion does not change if we use the $R_{AA}$ measurement from \cite{PHENIX_e} and ignore the $J/\psi$
feeddown contributions.
\begin{center}
\begin{figure}[t]
\resizebox{68mm}{!}{\includegraphics{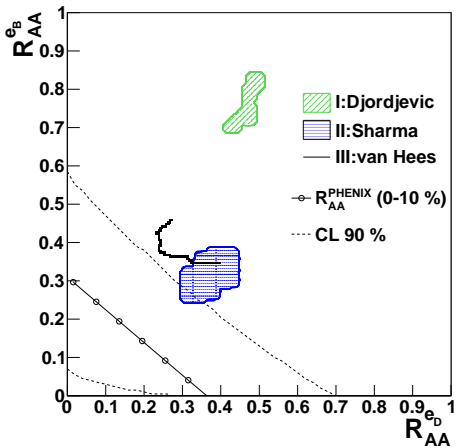}}
\caption{(color online)
Confidence level contours for nuclear modification factor $R_{AA}$ for electrons from $D$ ($R_{AA}^{e_D}$)
and $B$ ( $R_{AA}^{e_B}$) meson decays and determined 
by combining the $R_{AA}$ results and the $r_{B}$ measurement for $p_T >$  5 GeV/c. 
Three different models of $R_{AA}$ for $D$ and $B$ are described in the text.} 
\label{fig:fig4}
\end{figure}
\end{center}

For comparison, we also show model calculations in Fig.~\ref{fig:fig4}.
Model I includes radiative energy loss via a few hard scatterings with initial gluon density $dN_{g}/dy$ = 1000~\cite{eloss_1}.
Model I\hspace{-.1em}I includes cold nuclear matter effects, partonic energy loss and collisional dissociation \cite{eloss_dessociate}.
Model I\hspace{-.1em}I\hspace{-.1em}I assumes a large elastic scattering cross section 
associated with resonance states of $D$ and $B$ mesons in the QGP~\cite{eloss_reso}.
The model contours in Fig. \ref{fig:fig4} are calculated from the $p_{T}$ dependences of $R_{AA}$ for $D$ and $B$ decay in the interval 5 $<p_{T}\lesssim$  9 GeV/$c$.
For model I and I\hspace{-.1em}I, the uncertainties are also taken into account.
The experimental results are consistent with models I\hspace{-.1em}I  and I\hspace{-.1em}I\hspace{-.1em}I but are incompatible with model I.
Recently AdS/CFT theory has also been used to calculate the heavy quark energy loss in a strongly coupled quark-gluon plasma matter, for example \cite{heavy_AdsCFT1,heavy_AdsCFT2}. The theory also predicts strong suppressions for charm and bottom, and the ratio of the nuclear modification factors is proposed to differentiate AdS/CFT calculation from others \cite{eloss_AdsCFT1}.
 
In summary,  we measured the relative contribution from $B$ decays  to the non-photonic electron production in $p+p$ collisions 
at $\sqrt{s}=200$ GeV by using azimuthal correlations between non-photonic electrons and hadrons ($h$, $D^{0}$).
Our result indicates that the $B$ decay contribution increases with $p_{T}$ and is comparable to the contribution from $D$ meson decay  at $p_{T} \geq 5$ GeV/$c$.
Our measurement is  consistent with the FONLL calculation. 
The ratio of $N_{e_{B}}/(N_{e_D}+N_{e_{B}})$ combined with the large suppression of non-photonic electrons
indicates that $R_{AA}$ for electrons from $B$ hadron decays is significantly smaller than unity and therefore $B$ meson production is suppressed at high $p_{T}$ in heavy ion collisions.
The constraint on $R^{eD}_{AA}$ and $R^{eB}_{AA}$ will help to differentiate theoretical model calculations for heavy quark energy loss in the dense medium.

We thank the RHIC Operations Group and RCF at BNL, the NERSC Center at LBNL and the Open Science Grid consortium for providing resources and support. This work was supported in part by the Offices of NP and HEP within the U.S. DOE Office of Science, the U.S. NSF, the Sloan Foundation, the DFG cluster of excellence `Origin and Structure of the Universe' of Germany, CNRS/IN2P3, STFC and EPSRC of the United Kingdom, FAPESP CNPq of Brazil, Ministry of Ed. and Sci. of the Russian Federation, NNSFC, CAS, MoST, and MoE of China, GA and MSMT of the Czech Republic, FOM and NWO of the Netherlands, DAE, DST, and CSIR of India, Polish Ministry of Sci. and Higher Ed., Korea Research Foundation, Ministry of Sci., Ed. and Sports of the Rep. Of Croatia, Russian Ministry of Sci. and Tech, and RosAtom of Russia.


\end{document}